\documentclass{aa}
\usepackage{graphicx}
\begin{document}


   \title{Periodic orbits in warped disks}

   \author{Y. Revaz \and D. Pfenniger}

   \offprints{Y. Revaz}

   \institute{Geneva Observatory, University of Geneva, CH-1290 Sauverny, Switzerland\\
              email: Yves.Revaz@obs.unige.ch}

   \date{Received 22 November 2000/ Accepted 9 April 2001}

   \abstract{
   It is often assumed that a warped 
   galaxy can be modeled by a set of rings. 
   This paper verifies numerically
   the validity of this assumption by the study of periodic orbits populating a
   heavy self-gravitating warped disk.  
   The phase space structure of a warped model reveals that the circular periodic 
   orbits of a flat disk are transformed in quasi annular periodic orbits which conserve their stability. 
   This lets us also explore the problem 
   of the persistence of a large outer warp. In particular, 
   the consistency of its orbits with the density distribution is checked as a 
   function of the pattern speed.
   }
   
   \maketitle

%

\section{Introduction}

  In the conventional understanding of galaxies embedded in a kinematically hot 
  spheroidal or triaxial halos of dark matter, the frequently observed
  pronounced outer warps of neutral hydrogen, interstellar dust and, to a lesser
  noticeable extent, optical disks set a challenge for explaining their
  presumably long-lived existence, and particularly, their frequent
  quasi-straight line of nodes (LON), or pair of quasi-straight LON (Briggs\ \cite{briggs}).
  At least half the spirals possess a
  detectable warped HI and stellar disk(e.g., Briggs\ \cite{briggs}, 
  and more should remain
  undetected due to projection confusion.

  Hypotheses like 
  cosmic infall (Jang \& Binney\ \cite{Jang99}),
  gravita\-tional inter\-actions (Hern\-quist\ \cite{confwarp}; Wein\-berg\ \cite{weinberg95}, \cite{weinberg98}),
  normal modes (Sparke\ \cite{spar}; Sparke \& Caser\-tano\ \cite{case}),
  misaligned dark halos (Kuijken\ \cite{kuijken91}; Dubinski \& Kuijken\ \cite{dubi95}; Debattista \& Sellwood\ \cite{debattista99})
  or magnetic fields (Battaner et al.~\cite{batt}) have been proposed, 
  without clearly satisfying all the constraints provided  by the observations. See Binney (\cite{binney92}) 
  or Kuijken (\cite{kuijken00}) for reviews.

  But, as argued by 
  Arnaboldi et al.\ (\cite{arnaboldi}),
  Becquaert \& Combes\ (\cite{becq}) and
  Reshetnikov \& Combes\ (\cite{resh}), also in the context of polar rings,
  the evidence for 
  non-self-gravitating HI disks
  is actually weak.  Instead, the opposite assumption of
  coexistence of warps containing most of the dynamical mass is not in
  contradiction with the available observations.  Such an assumption of
  thick and flaring disks of visible and dark matter in the form of cold
  gas approximately proportional to neutral hydrogen has been proposed by Pfenniger et
  al.\ (\cite{pfenn94}) to account for many more known facts about spirals.
  
  For the warp problem, the self-gravitating disk assumption is attractive for several reasons.  
  HI observations reveal that 
  most of the galaxies possess an inner flat disk, while the warp develops beyond a   
  specific radius (Briggs\ \cite{briggs}; Burton\ \cite{burton}). 
  Such a comportment, which is the signature of a radical change of the dynamic along the radius, can hardly be 
  explained by a hot halo which would impose a uniform dynamic over a larger scale. 
  In contrast, a strong rotational support in the dark matter component offers a 
  natural explanation for 
  the straight lines of nodes.  
  They would result as a natural consequence of the weak diffusive 
  properties of angular momentum in a self-gravitating disk:
  if, following for example an accretion event,
  angular momentum is deposited in a self-gravitating disk, bending
  its outer parts, the angular momentum will {\it slowly\/} diffuse
  across the whole disk in a monotonous way, because angular momentum is
  a quasi conserved quantity.  As its diffusion is slow, the line
  of nodes must be quasi-straight and persist for many rotational
  periods.
  Finally, self-gravitating optical disks are in agreement with the often stated maximum disk 
  interpretation of observations of the Milky Way and other spirals (see for example
  Gerhard\ \cite{gerhard00}).
  
  In numerous models of warped galaxies, and also polar ring models, stars are assumed to move
  along circular rings which are tilted as a function of radius. Because stars are driven by
  periodic orbits, which are the backbone of galaxies, such models implicitly 
  presuppose  the existence of
  stable circular tilted periodic orbits. Yet, to our knowledge they have never been verified, 
  and do not appear obvious in non-spherical geometries. It is far from 
  obvious that a perturbation in the form of a warp will conserve the well known circular
  orbits of a  flat disk. In order to verify the latter assumptions, it is useful to 
  find the exact periodic orbits existing in a warped disk. 
      
  Moreover, understanding the stable  periodic orbits allows us to grasp the basic properties of the other
  quasi-periodic  orbits in an efficient way, particularly when no analytical tools exist. 
  Such an approach has been very useful in the context of barred
  galaxies for which the complexity of motion is substantial (e.g., Contopoulos \&
  Papayannopoulos\ \cite{conto};  Pfenniger\ \cite{pfenn84}).
  
  For studying the principal periodic orbits in a galactic potential, 
  even rough representations
  of the potential are sufficient to yield the basic properties of the 
  main periodic orbits, as that shown in numerous situations. 
  In this work, the disk autogravitation will play a dominant role.
  Therefore, the following results will be applicable as long as the local density 
  is dominated by the disk. 

  The alternative case would be the existence of a hot dark matter halo. 
  In this case, one must admit a center dominated by the halo density,
  because the density of a warm gravitating system increases at the center. 
  The following implications should be explored:
  
  \begin{itemize}

    \item[1)] A hot halo is aligned with the central stellar disk. The outer 
  gaseous disk acts as a tracer of the main periodic orbits in the halo. 
  The only reason for the disk to warp is then to follow the orbits generated by a  
  vertical 1:1 resonance. 
  But in spheroidal halos, stable warped orbits do not exist due to the invariance 
  in azimuth of the potential.  
  To avoid this problem, one has to take in account both the mass of the halo and 
  the self-gravity of the disk (Sparke\ \cite{spar}; Sparke \& Casertano\ \cite{case}).  
  In triaxial halos more periodic 1:1 orbit families exist.  
  1:1 resonances are strong only in rapidly rotating triaxial potentials 
  (e.g. Mulder\ \cite{mulder83}), and associated orbits would be compatible 
   with warps only outside the corotation radius: the dynamics would resemble the one 
of a rescaled triaxial bar (Pfenniger\ \cite{pfenn84}).  But such triaxial halos appear difficult
  to reconcile 
   with the axisymmetrization that follows the inclusion of even a weak 
   fraction of dissipative matter (Dubinski\ \cite{dubi94}). 

    \item[2)] The gaseous disk is aligned with the outer part of a hot halo, and
  the optical disk is tilted.
  This case appears dynamically highly unstable because the optical disks appear 
  maximum (cf. Gerhard 2000), i.e., self-gravitating to a large extent.
  
    \item[3)] The halo follows the warp of the disk.
  Such a halo can survive if it possesses enough angular momentum and therefore 
  should be flattened.
  This case becomes very similar to the case of the massive disk,
  and the comportment of the periodic orbits is not changed dramatically.

  \end{itemize}
  
  In this paper we
  present succinctly a study of periodic orbits in a simple model of
  warped disk galaxies, in which the
  stability and consistency with the density distribution is also 
  determined. The influence of a global rotation of the warp pattern is
  examined too. A subsequent study will investigate the orbits in self-consistent 
  N-body models of warps.


\section{Model}


\subsection{Warped disk model}

In order to compute periodic orbits, we have based our model on a superposition
of three simple Miyamoto-Nagai potentials (Nagai \& Miyamoto\ \cite{nagai}). 
The three components can
be viewed as representing respectively a bulge, a visible disk and a gas disk 
containing a large amount of dark matter.
In cylindrical coordinates, the total potential can be written as :
        \begin{equation}
        \label{pot_0}
        \Phi_0(R,\phi,z) = -\sum_{i=1}^{3} \frac{GM_i}
             {\sqrt{R^2+\left(a_i+\sqrt{z^2 + b_i^2}\right)^2}}\ .
	\end{equation}
Observations suggest that warps seen on edge take the shape of an 
integral sign, i.e., the
disk deformation is proportional to the cosine azimuthal 
angle $\phi$ and increases with radius.
Analytically, the warp is well represented by a term $\Delta z$ where :
        \begin{equation}
        \label{delta_z}
        \Delta z = w R^2 \cos\phi,
        \end{equation}
and $w$ is an adjustable parameter characterizing the warp's amplitude.

Replacing $z$ by  $z-\Delta z$ in the potential (\ref{pot_0}) we get an analytical
warped disk potential model :
        \begin{equation}
        \label{pot_w}
        \Phi_{w}(R,\phi,z) = -\sum_{i=1}^{3} \frac{GM_i}
             {\sqrt{R^2+(a_i+d_i)^2}},
	\end{equation}
with
        \begin{equation}
        \label{d_i}
        d_i=\sqrt{\left(z-w R^2 \cos\phi\right)^2 + b_i^2}\ .
	\end{equation}
Note that the potential is 
bi-symmetric with respect to the $y=0$ plane, $\Phi(x,y,z) =
\Phi(-x,y,-z)$.  

\subsection{Parameters}

\begin{figure*}
\resizebox{\hsize}{!}{\includegraphics[angle=-90]{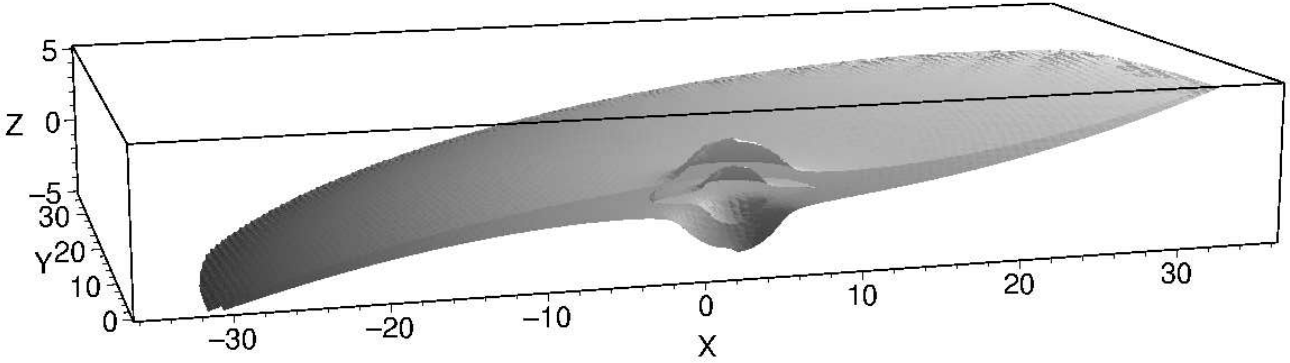}}
\caption{3D isodensity surface of the warped model at $5\cdot10^{-4}$ 
        and
	$5\cdot10^{-5}\textrm{M}_{\odot}\cdot \textrm{kpc}^{-3}$
	($w=0.005$~kpc$^{-1}$). The surface is sliced with respect to the $y=0$ plane.}
\label{dens_xz_m1w}
\end{figure*}

The parameters $a_i$, $b_i$ and $GM_i$ have been chosen 
to provide a typical
rotation curve, increasing linearly below $3\, \rm kpc$ and staying flat up to  
$40\, \rm kpc$, beyond which it decreases smoothly.
For the bulge, $a_b=0$, $b_b=1.5$ and $GM_b=0.066$, which is equivalent to a Plummer
sphere.
The parameters for the visible and gas disks are respectively~: $a_d=6.5$,  $b_d=0.5$, 
$GM_d=0.171$, $a_g=25$, $b_g=0.5$, $GM_g=0.763$.
For convenience the length unit is chosen to be the kpc and time unit the Myr. Fixing
the gravitational constant $G$ to a value of 1, the mass unit corresponds to 
$2.2 \cdot 10^{11} \textrm{M}_\odot$. With this set, the velocity unit is about
$1000\, \textrm{km} \cdot \textrm{s}^{-1}$.
Fig.~\ref{dens_xz_m1w} shows the  isodensity 
curves corresponding to the model. Well above  the galaxy disk, negative density regions do
exist, however our results are not
spoiled by negative densities as long as orbits do not cross such regions.
The amplitude of the deformation $w$ is chosen to be equal to 
$0.005\,\textrm{kpc}^{-1}$, developing a warp of the same order as the one of the Galaxy. 
For example, see the
values indicated by Burton\ (\cite{burton}) or Smart et al.\ (\cite{smart}).
Hence, the results obtained are applicable in typical cases of observed warped galaxies.

\subsection{Hamiltonian}

In the present model, we allow the possibility for a global rotation of the
warped potential about the $z$-axis.  This rotation is parametrised by the angular
pattern speed  $\Omega_p$ or, equivalently, by the corotating radius
$R_c$, taking a negative value for retrograde rotation. Thus, the Hamiltonian 
in the rotating frame of reference with angular speed $\Omega_p$ is :
        \begin{equation}
        \label{hamilt}
        H=\frac{1}{2}\left( p_x^2 + p_y^2 + p_z^2 \right)
	+\Phi(x,y,z)-\Omega_p\left( x\,p_y-y\,p_x \right),
	\end{equation}
where the variables $p_x$, $p_y$, and $p_z$ are the respective canonical 
momenta of $x$, $y$, 
and $z$. The equations of motion to integrate are :
	\begin{equation}
	\label{motion}
	\begin{array}{l c l c l c l}
	\dot x&=&p_x + \Omega_p y,& & \dot p_x &=& -\partial_x\Phi_w + \Omega_p p_y,\nonumber\\
	\dot y&=&p_y - \Omega_p x,& & \dot p_y &=& -\partial_y\Phi_w - \Omega_p p_x,\\
	\dot z&=&p_z,		& & \dot p_z   &=& -\partial_z\Phi_w,\nonumber
	\end{array}
	\end{equation}
with $\Phi_w$ given by Eq.~(\ref{pot_w}).

\subsection{Numerical method}

The periodic orbits are found by numerically determining the fixed points
of the 4D Poincar\'e map ($T$) at $y=0$, $\dot x <0$ generated by the
equations of motion (see, e.g., Pfenniger \& Friedli\ \cite{pfenn93}).  
The algorithm uses in particular the method
proposed by H\'enon\ (\cite{henon}) giving the intersection between an orbit
and a surface, and a least squares stabilized Newton-Raphson root-finding 
procedure. 

The stability of the orbits are determined by the eigenvalues of 
the Jacobian of the Poincar\'e map ($\nabla T$). 
Since  the system is Hamiltonian and the motion is described by 
real numbers, 
the four eigenvalues of $\nabla T$ occur by conjugate and inverse pairs
and it is possible
to condense the information with two stability indexes $b_1$, $b_2$:
        \begin{equation}
        \label{stabindex}
        b_i  =  -(\lambda_i + \lambda_i^{-1}), \quad i=1,2, \\
	\end{equation}
where $\lambda_1$ and $\lambda_2$ represent a pair of reciprocal eigenvalues.
A periodic orbit is stable only when $b_1$ and $b_2$ are real and
$|b_1|$,$|b_2|<2$. It is unstable in all other cases. 
If $|b_1|=2$ or $|b_2|=2$, or if $|b_1|=|b_2|$, 
at least two eigenvalues are equal, 
$\nabla T$ is degenerate and eventually allows a bifurcation. 
For a more complete description of the instability cases,
see Pfenniger \& Friedli\ (\cite{pfenn93}).

In this work, we will focus our investigations on orbits following the potential
disk. Hence, we choose initial conditions with a starting position on the line 
of nodes, here equivalent to the $y$-axis. 
We concentrate on the main orbit families keeping the symmetry of the potential, thus 
the initial velocity is perpendicular to it ($p_y(0)=0$) and points to arbitrary
negative values of $x$ ($p_x(0)<0$).  
In this way, the
free parameters in the initial conditions are the Jacobi constant ($\approx$
energy, $H$), the position $y(0)$ along the $y$ axis, and the velocity component 
$p_z(0)$ 
along the $z$
direction. This latter parameter will be small compared to $p_x(0)$, which  avoids
a situation where orbits stray too far from
the disk and penetrate negative density regions. 
This point has been systematically checked because, a priori orbits may explore 
regions far from the expected ones.


\section{Families of periodic orbits}


\subsection{Families without perturbation}

In order to understand the influence of the warp, we
first look at the model without deformation ($w=0$) which is completely
axisymmetric. The rotation of the potential is set to zero ($\Omega_p =0$).

\begin{figure}
\resizebox{\hsize}{!}{\includegraphics{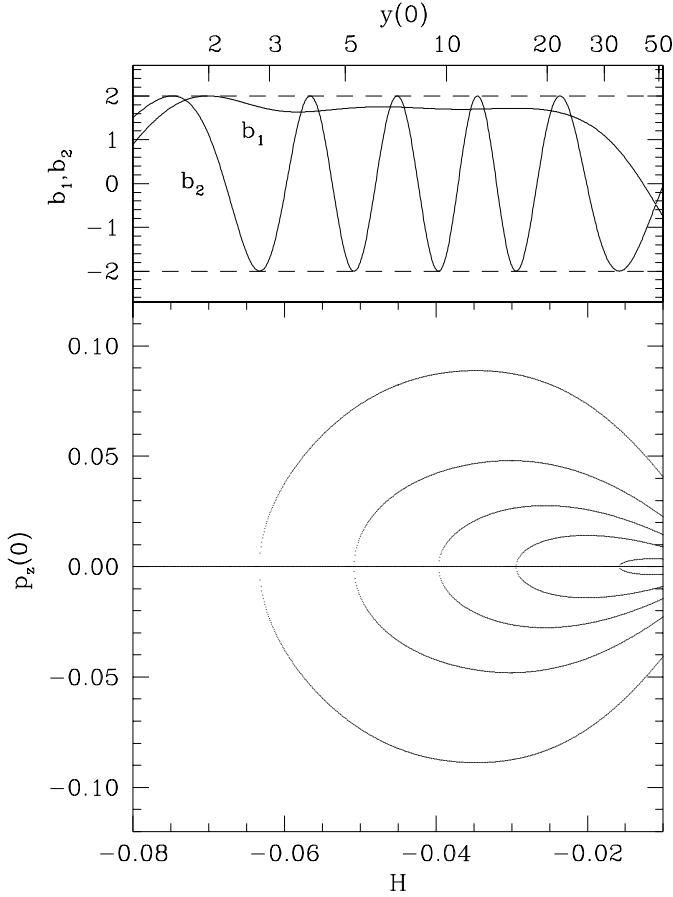}}
\caption{$H$-$p_z(0)$ phase space for $w=0$.}
\label{pz_h_0}
\end{figure}

Below the orbit, families are shown with their initial starting point in the 
$H-p_z(0)$ diagram.
In Fig.~\ref{pz_h_0}, bottom, we
recognize the circular orbit family (horizontal line at $p_z(0)=0$). 
The stability indices for this 
family are traced at the top of Fig~\ref{pz_h_0}. In this particular case, $b_1$ 
indicates the stability in
the galaxy plane while $b_2$ corresponds to the stability transverse to it. 
Both indices remain in the interval $[-2,+2]$, insuring the stability for this
family.

At $H=-0.0701$, $b_1=+2$, $\nabla T$ is also degenerate 
with two eigenvalues equal to $-1$. A bifurcation occurs through period doubling.
This bifurcation coincides with the resonance between the radial frequency
$\kappa$ and the circular frequency $\Omega$ ($2\kappa=3\,\Omega$).

At $H=-0.0748$, $-0.0566$, $-0.0451$, $-0.0345$, and $-0.0236$, 
$b_2=+2$ generating transverse
bifurcations through period doubling. 
These bifurcations coincide with the resonances between the transverse frequency
$\nu$ and the circular frequency $\Omega$ ($2\nu=(2k+1)\,\Omega$, $k=1,\ldots 5$). 
Orbit families from period doubling bifurcations will not 
be discussed further in this paper.

Transverse bifurcations keeping the same period occur
at $H=-0.0632$, $-0.0508$, $-0.0396$, $-0.0294$, and $-0.0158$. 
In this case $b_2=-2$ and 
$\nu = k\, \Omega$, $k=2,\ldots 6$.
The two latter families have a non zero $p_z(0)$
initial velocity and also oscillate round the $z=0$ plane crossing it
respectively $2k$ and $2(k+1)$ times per period.
All these sub-families are marginally stable with $b_2=-2$ due to the 
axisymmetry of the potential.

%

\subsection{Influence of the warp}

The break in the $z$ symmetry transforms the family of circular orbits in the unperturbed 
model into a set of families $pk$ (main families) with $p_z(0)\neq 0$. 
The significance of the indice $k$ will be explained
further. 
Except at bifurcations, the $pk$ families are stable.

The shape of the corresponding orbits can be approximated
with the following parametric function:
        \begin{eqnarray}      
	R(H) & = & R_0(H) + A_R(H) \cos(2\phi - \pi), \\
	z(H) & = & z_0(H) + A_z(H) \cos\phi,   
	\label{orbit_shape}
	\end{eqnarray}
parametrised by the azimuthal angle $\phi$. For increasing $H$, $A_R$ increases from
0 to $1\,$kpc, while $A_z$ increases from 0 to $8\,$kpc. 
This latter variation corresponds
almost exactly to the amplitude of the warp as a function of radius 
($A_z \cong wR^2$).  In other words, the main orbit family follows
the density maximum at any radius.

\begin{figure*}
\centering
\resizebox{15cm}{!}{\includegraphics{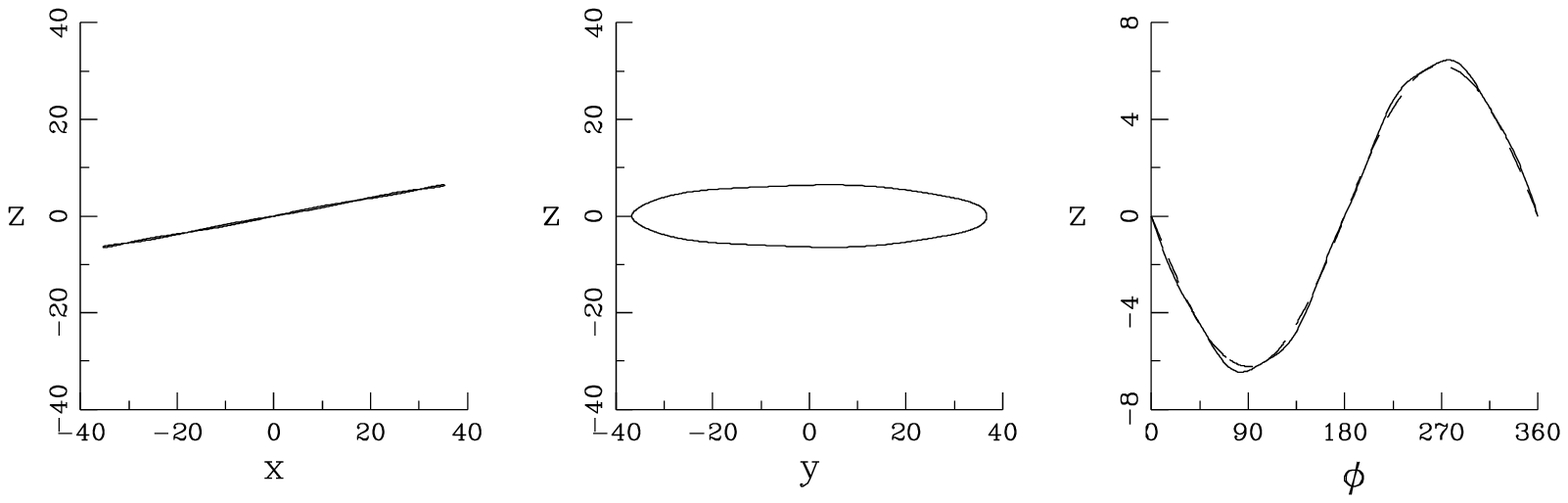}}
\resizebox{15cm}{!}{\includegraphics{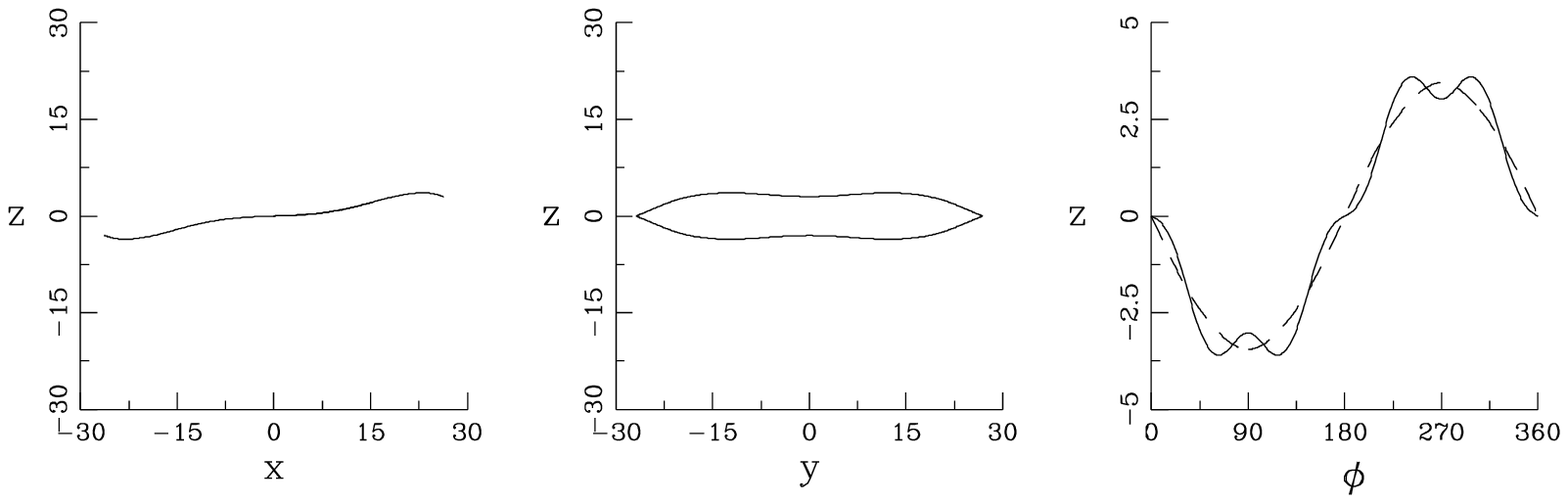}}
\resizebox{15cm}{!}{\includegraphics{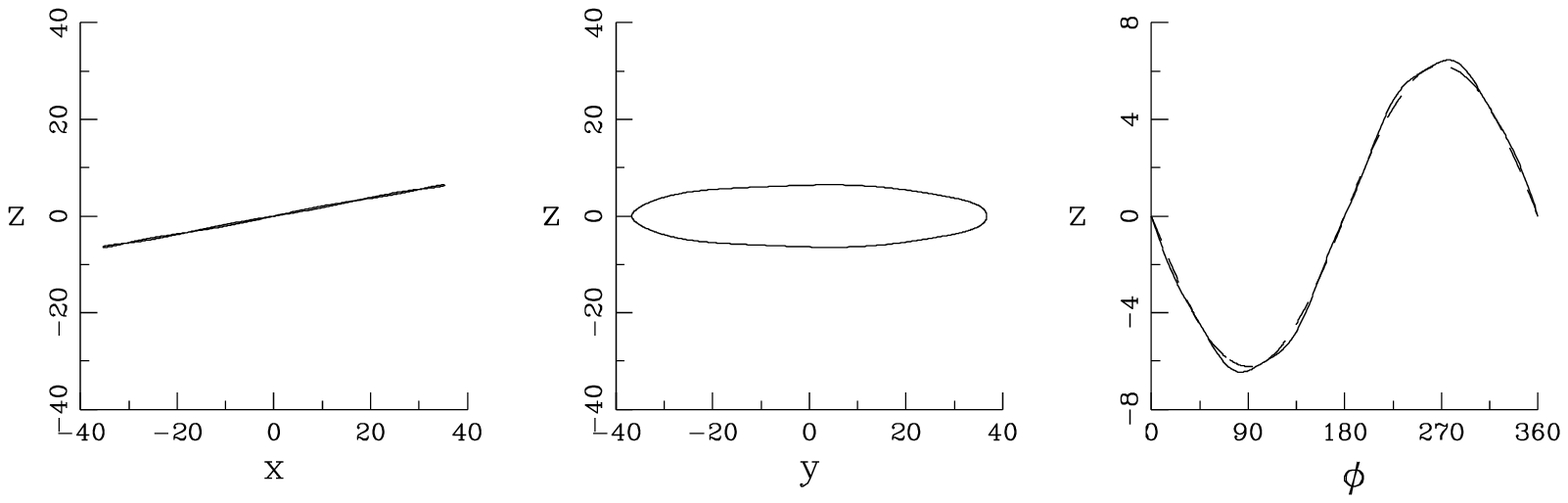}}
\caption{Projections over the $y=0$, $z=0$ planes and
the evolution of $z$ as a function of azimuthal angle $\phi$ of orbits 
of the families $s4+$ ($H=-0.03$), $s5+$ ($H=-0.02$) and $s6-$ ($H=-0.015$). 
The dashed line represents the
potential minimum of the warped disk at a fixed radius which is similar to the main
family.}
\label{projection}
\end{figure*}

Another influence of the warp is to move the bifurcations
toward slightly lower energies. The corresponding transverse bifurcations 
arise at $H=-0.0633$, $-0.0509$, $-0.0398$, $-0.0298$, and $-0.0176$.
Moreover, a new bifurcation  takes place within the limits of the diagram at $H=-0.0108$,
where the subfamily starting at $H=-0.0176$ rejoins the main family. The other
families would also rejoin the main family at higher energies, yet  they would reach
radii larger than $y(0)=50\,\rm kpc$, which was chosen as a limit for our study of a
warped disk. 
\begin{figure}
\resizebox{\hsize}{!}{\includegraphics{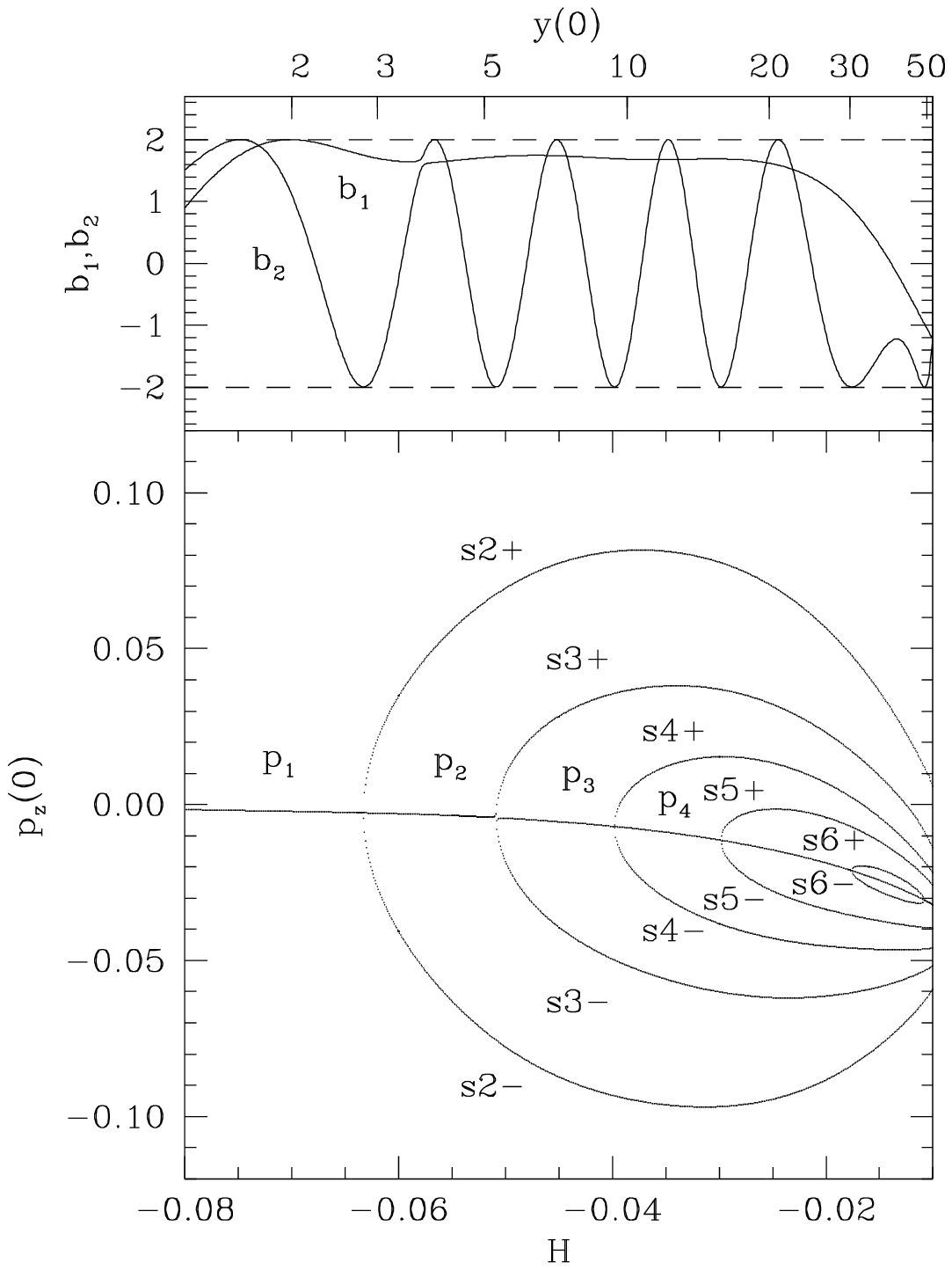}}
\caption{$H$-$p_z(0)$ phase space ($w=0.005$).}
\label{pz_h}
\end{figure}
The subfamilies coming from
the corresponding transverse bifurcations are given the symbol 
``$sk\pm$'' where $k$ gives the frequency
ratio $\nu / \Omega$ at the corresponding bifurcation and the sign is the
one of the difference of $p_z(0)$ between the subfamily and the main family.
These families evolve in the same way as family $pk$, i.e. they follow the warped 
disk, so may be significantly populated in a real disk. Note
that the families with an odd $k$ conserve the bi-symmetry of the potential, which is not the
case when $k$ is even.
Fig.~\ref{projection} displays a set of three orbits extracted from different subfamilies,
$s4+$, $s5+$ and $s6-$. The right panels show the elevation $z$ of the orbits as a function
of the azimuth $\phi$ in comparison to the elevation of the corresponding main family
(dashed lines). The more the energy grows, the more the subfamilies follow well the main
orbit.
Although in the absence of the warp all subfamilies are stable, this is no longer the case 
with a weak warp. This point will be discussed further below.

\subsection{Surfaces of section around bifurcations}

\begin{figure}
\resizebox{\hsize}{!}{\includegraphics{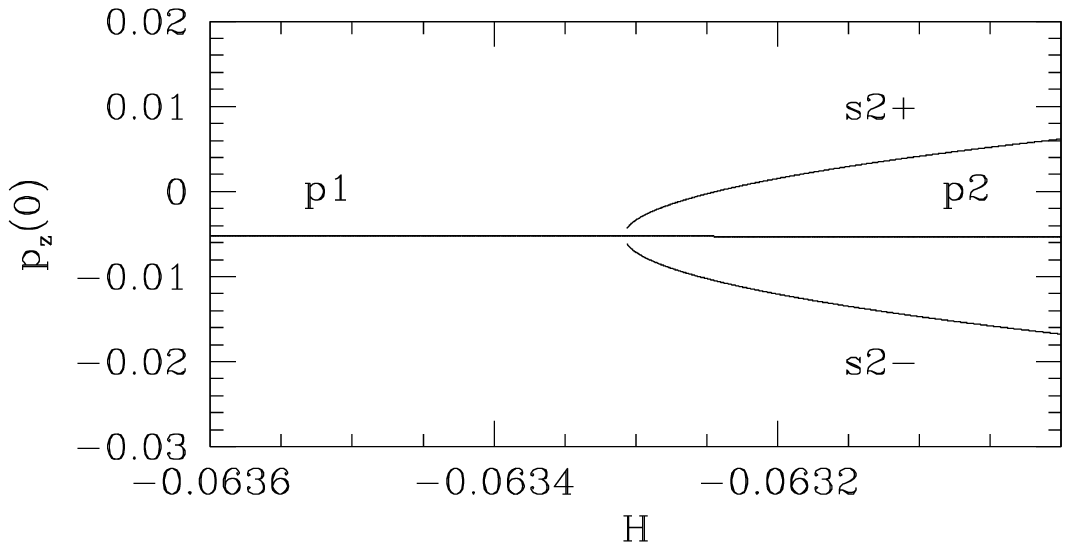}}
\caption{$H$-$p_z(0)$ and $H$-$y(0)$ phase space around the first bifurcation ($w=0.01$).}
\label{r1}
\end{figure}

\begin{figure}
\resizebox{\hsize}{!}{\includegraphics{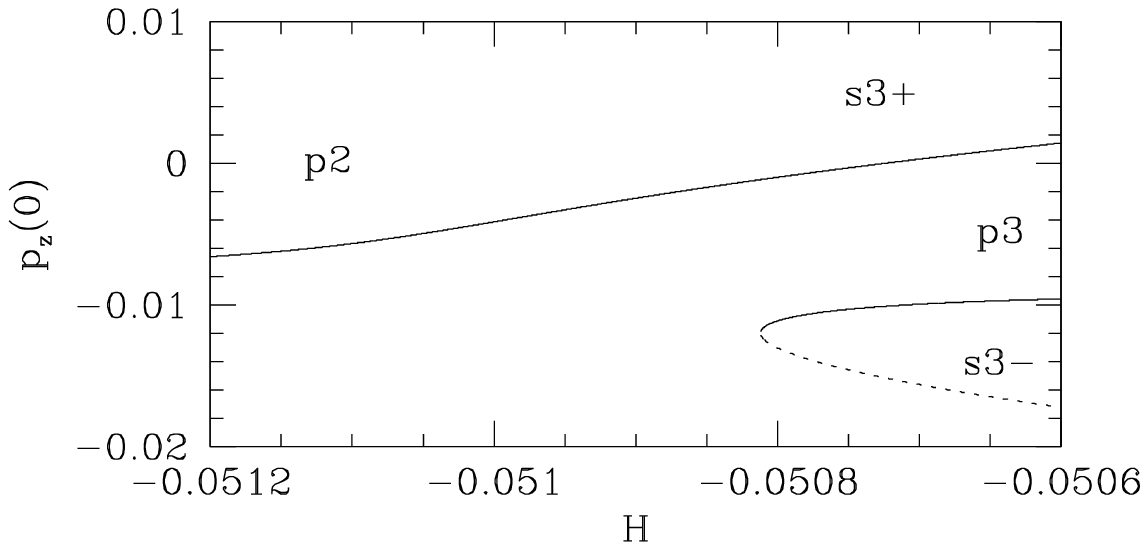}}
\caption{$H$-$p_z(0)$ and $H$-$y(0)$ phase space around the second bifurcation ($w=0.01$).}
\label{r2}
\end{figure}

If the warp has only a slight influence on the shape of the orbits, 
it is important to look carefully at the behavior of the bifurcations in presence of 
the perturbation.
Fig.~\ref{r1} and Fig.~\ref{r2} present a magnification of the $H$-$p_z(0)$ phase space around the 
first ($r1$) and second ($r2$) bifurcations. 
Here the periodic orbits are computed with $w=0.01$ in order to increase the effect of 
the warp. The physical correspondent would be a galaxy with a warp twice as high as that
observed in the Milky Way at a radius of $30\,\textrm{kpc}$.

\begin{figure*}
\resizebox{\hsize}{!}{\includegraphics{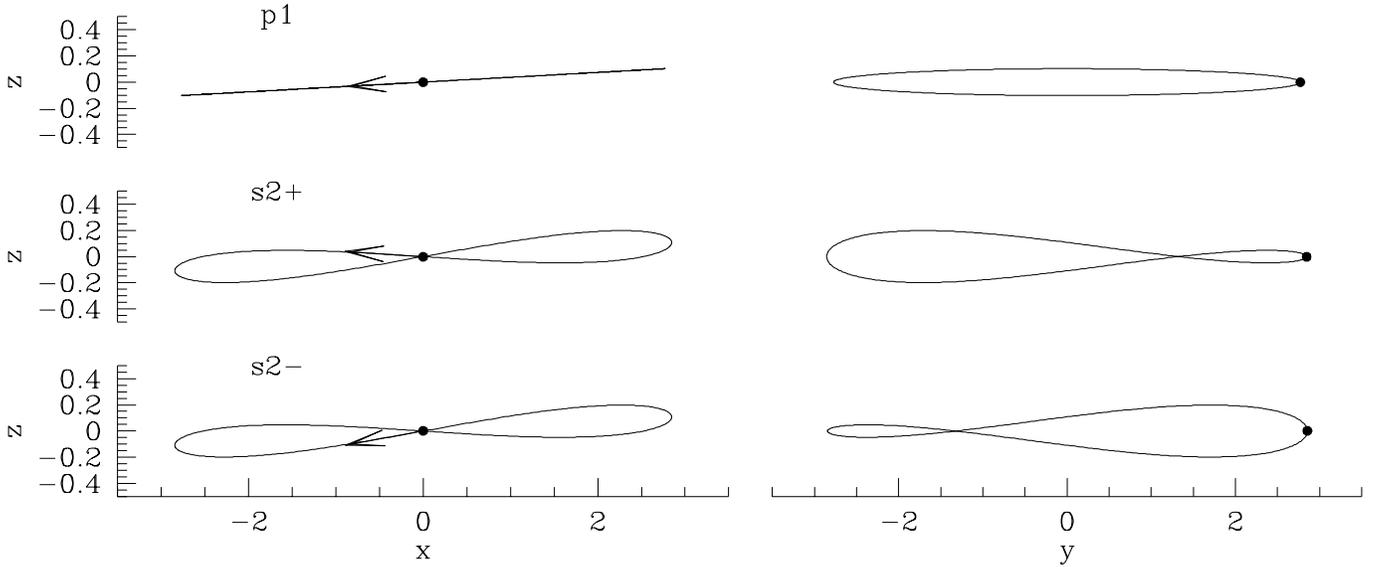}}
\caption{Shape of three orbits linked to the $r1$ bifurcation ($H=-0.06$). 
The dot indicates the starting point of the computation 
and the arrow shows the direction of the corresponding initial
velocity.}
\label{orbit_r1}
\end{figure*}

Except for the weak decreasing of $p_z(0)$ with respect to $H$, the type of the first 
bifurcation (around $H=-0.0633$) remains
similar to the one without perturbation (pitchfork). The two new generated families 
($s2+$ and $s2-$) are both stable. 
They appear at the same energy as two other unstable
families, namely $z2+$ and $z2-$. 
These are not seen in Fig.~\ref{r1} because their projection
merges with the $p2$ family. They are distinguished from it by a non zero $z(0)$ and $p_y(0)$.
Projections of families $s2+$ and $s2-$ are plotted in Fig.~\ref{orbit_r1}.

\begin{figure}
\resizebox{\hsize}{!}{\includegraphics{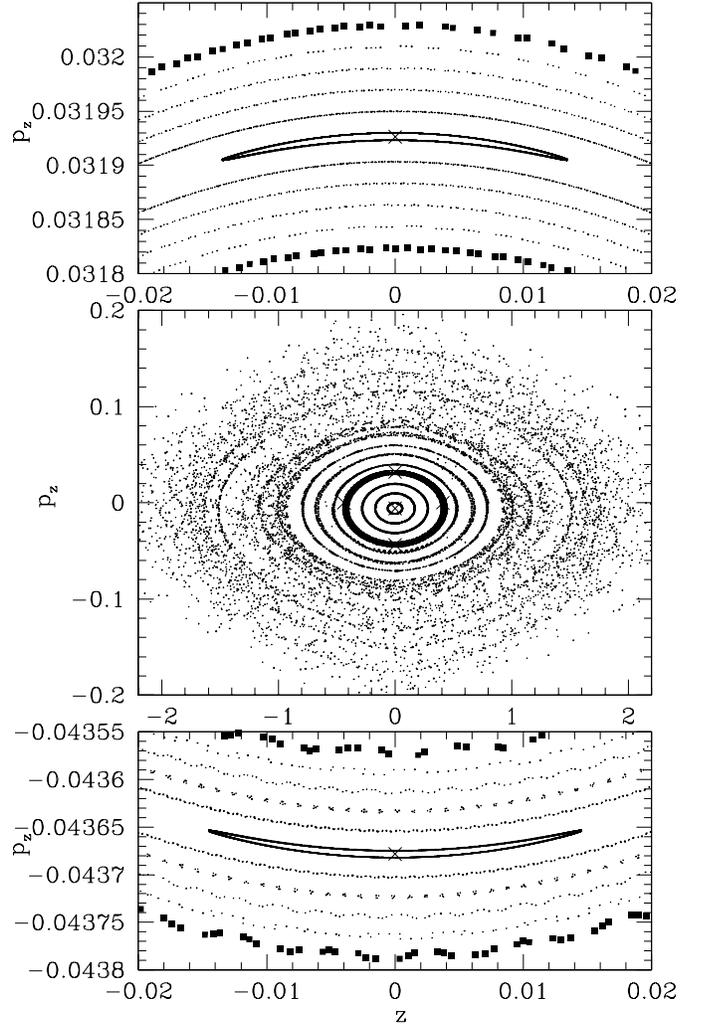}}
\caption{Section $(z,p_z)$ at $H=-0.06$, near the bifurcation $r1$. The crosses
represent the three stable periodic families $s2+$ (top), $p2$ (middle) and $s2-$ 
(bottom) and the two unstable $sz+$ (right) and $sz-$ (left). 
The upper and lower panels show a
zoom of the section around $s2+$ and $s2-$.  In this  panels the
separatrix is marked with squares while in the middle it is marked with a bold line. The
crosses indicate the position of the periodic families.}
\label{r1susec}
\end{figure}

In order to estimate the importance of orbits associated with a stable periodic orbit, we have
computed the surface of the section $(z,p_z)$ near the bifurcation ($H=-0.06$). Moreover, it
allows us to know the number of effective integrals of motion in this specific region of phase space.
In Fig.~\ref{r1susec} the stable $p2$ family is represented by the cross near the center
($z=0$, $p_z=-0.0058$) and is surrounded by quasi periodic families. The thinness of these
curves indicates that the motion is well decoupled in $z$ and a third integral exists. 
At $p_z=0.031,z=0$ and $p_z=-0.043,z=0$  we find respectively the $s2+$ and $s2-$ 
families. 
The upper and lower panels show details of the section around these two families. Very thin 
stability islands exist. For comparison, the width in $p_z$ 
of the separatrix (squares points) is about $1/100$ times smaller than the width between
families $s2+$ and $s2-$. 
This gives an indication of the phase space occupied by orbits associated with
these families.
The unstable $z2+$ and $z2-$ families appear on the separatrix, 
at $p_z=-0.0054,z=0.427$ and $p_z=-0.0054,z=-0.427$.
Outside the separatrix, the invariant curves are regular until $p_z= 0.06$ where 
the motion in $z$ begins to be strongly coupled to the motion in $x$ and $y$. 

\begin{figure}
\resizebox{\hsize}{!}{\includegraphics{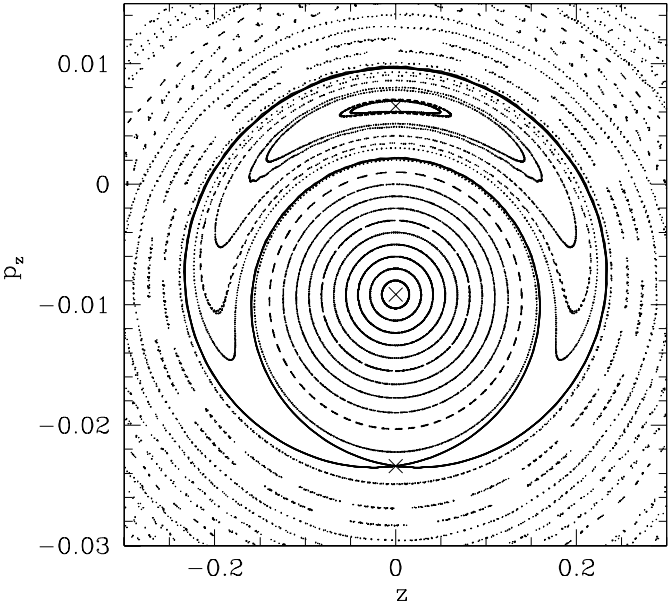}}
\caption{Section $(z,p_z)$ at $H=-0.05$, after the bifurcation $r2$. 
The separatrix is marked with a bold line. The crosses indicate the position of 
the three periodic families $s3+$ stable (top), $p3$ stable (middle) and $s3-$ 
unstable (bottom)}
\label{r2_2susec}
\end{figure}

\begin{figure*}
\resizebox{\hsize}{!}{\includegraphics{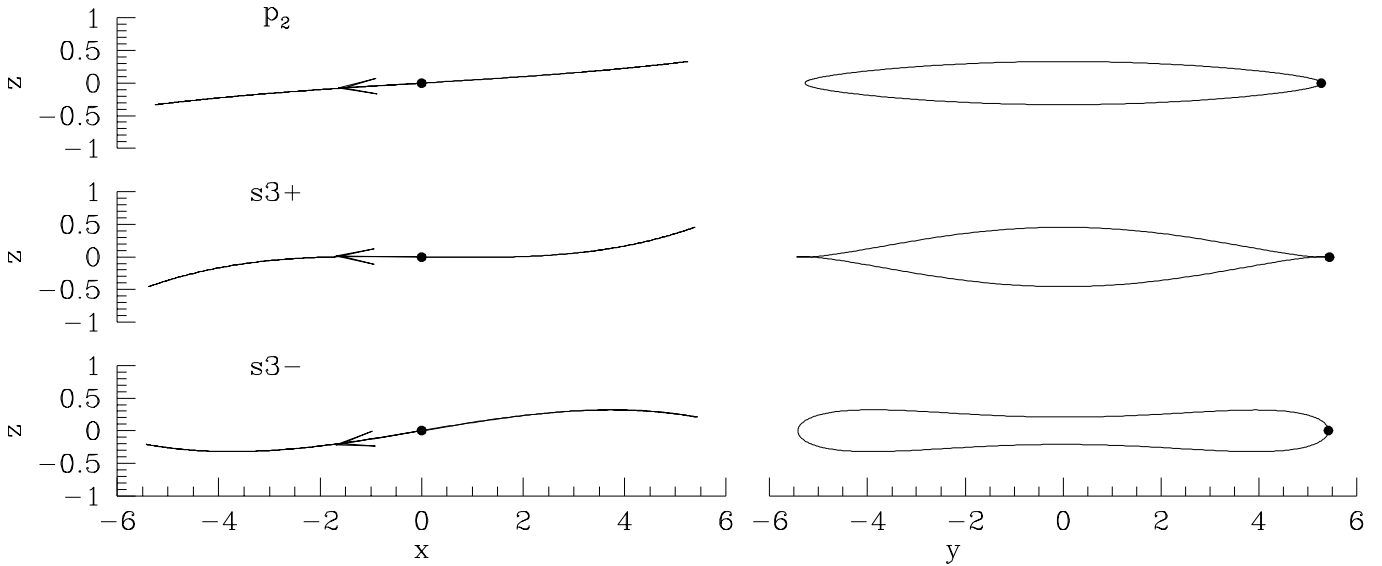}}
\caption{Shape of three among the four orbits linked to the $r2$ bifurcation 
($H=-0.0509$).
The dot indicates the starting point of the computation 
and the arrow shows the direction of the corresponding initial
velocity.}
\label{orbit_r2}
\end{figure*}

The behavior of the second bifurcation $r2$ around $H=-0.0509$ is quite different. 
Fig.~\ref{r2} reveals that the family $s3+$ is in fact the extension of the main family 
$p2$. This arises because the perturbation favors a higher $p_z$. 
The $s3+$ family is also stable. 
At $H=-0.0508$ the families $p3$ and $s3-$ are created simultaneously. 
They are respectively stable and unstable. The whole bifurcation forms a pitchfork
with a symmetry breaking. 
The shapes of three of the four families involved in this bifurcation are 
presented in Fig.~\ref{orbit_r2}.
The section after the bifurcations is shown in Fig.~\ref{r2_2susec} ($H=-0.05$). 
The stable $p3$ family appears at $p_z=-0.009$ and is surrounded by a set of quasi 
periodic orbits embedded in the separatrix. 
The unstable $s3-$ family belongs to the separatrix at $p_z=-0.023$. 
The $s3+$ family is located at $p_z=0.006$ and its associated
orbits occupy the crescent-shaped region between the separatrix.
Outside, the invariant curves are regular until $p_z=-0.03$ where the motion in $z$ is
not longer decoupled.
Looking at Fig.~\ref{r1susec} and \ref{r2_2susec}, 
one can expect a chaotic region around the separatrix.  
Due to the thinness of the region around the unstable points, 
it is very difficult to find it numerically.

Our study of the bifurcation is limited to the two first bifurcations, $r1$ and $r2$. 
The behavior of the following bifurcations are similar to the first two. The
$rk$ bifurcation with an odd (resp. even) $k$ is of the same type as $r1$ (resp. $r2$). 
This difference is related to the symmetry of the families which depends on the parity of
$k$. 

In summary, the warp does not destroy the strength of the main
family, which at any $H$ possesses a comfortable surrounding stable region. 
For an odd $k$, the effect of the warp is :
(i) to change the stability of
$sk-$ subfamilies and 
(ii) to increase the phase space allowed to the associated $sk+$
families. For an even $k$, the subfamilies are quasi unchanged, preserving a very 
thin phase space around the stable subfamilies $sk\pm$.

\subsection{Influence of rotation}

To study the influence of the rotation, we first look for
Lagrangian points. 
The latter are found by setting the right terms of equations~(\ref{motion}) to zero.
Substituting $p_x$, $p_y$ and $p_z$, the Lagrangian points are solutions of the following equations :
	\begin{eqnarray}
	\label{lagranges}
	\partial_x\Phi_w & = & \Omega_p^2 x,\nonumber\\
	\partial_y\Phi_w & = & \Omega_p^2 y, \\
	\partial_z\Phi_w & = & 0.\nonumber
	\end{eqnarray}
With the form (\ref{pot_w}) of the warped potential, one can easily see that the 
$z$ coordinate must satisfy :
	\begin{equation}
	\label{zcond_lagranges}
	z = \Delta z,
	\end{equation}
with $\Delta z$ defined in equation~(\ref{delta_z}). This causes the two first equations 
of~(\ref{lagranges}) to become :
	\begin{eqnarray}
	\label{lagranges_2}
	\partial_x\Phi_0 = \Omega_p^2 x,\\
	\partial_y\Phi_0 = \Omega_p^2 y, \nonumber
	\end{eqnarray}
which corresponds to the definition of the corotation, where $\Phi_0$ is the potential
without perturbation (Eq.~(\ref{pot_0})). 
Thus the Lagrangian points are degenerated, forming an annulus following 
the density maximum with a projection on 
the $z=0$ plane corresponding to the corotation without perturbation ($w=0$). 

Despite the fact that a direct and retrograde rotations respectively add and remove
resonances, the type of bifurcations as well as the stability and the shape of orbits are 
not affected by a global pattern speed. 
This has been tested in the range $-30 > R_c$ and $R_c > 30\,\textrm{kpc}$, where $R_c$ 
is the corotation radius and a negative value corresponds to a retrograde rotation.

For a radius less than $40 \textrm{kpc}$, both prograde and retrograde 1:1 
resonances are not observable in this range of pattern speed.
For a corotation of $7\,\textrm{kpc}$ they appear only beyond  
$40\,\textrm{kpc}$ (resp. $30\,\textrm{kpc}$). Thus they have not been studied.

The main influence appears when we look at the consistency of orbits with the mass
density. If the warp is mostly self-gravitating and made of thin and distinct tube
orbits, one can check the self-consistency constraint by noting that the
spatial occupation of a periodic orbit is locally inversely proportional to its
local speed ($t_{so} \sim 1/|v|$). 
The reason is explained by the fact that at a given point of the orbit, the
local speed remains constant in time. This argument is not available at exceptional points, 
for example the points where an orbit crosses itself.
Since the density is proportional to the spatial occupation time, it must also
be proportional to the inverse speed along the orbit
($\rho \sim t_{so}\sim 1/|v|$).  
Strictly this condition is only fulfilled
by a structure entirely made of distinct exactly periodic orbits, such
as a disk made of circular orbits.  Nevertheless the check is useful
in this problem because few hot orbits far from periodic round orbits
are expected to exist.

In practice, the consistency has been calculated using the indice $I$ defined 
by the norm :
	\begin{equation}
	I(\Omega_p,H)=\frac{1}{2\pi}\left[\int_0^{2\pi}\left[\Delta\rho(\phi)-
	\Delta u(\phi)\right]^2d\phi\right]^{1/2},
	\label{cons}
	\end{equation}
with
	\begin{equation}
	\Delta\rho(\phi)=\frac{\rho(\phi)-\rho_{min}}{\rho_{max}-\rho_{min}},
	\label{rhostar}
	\end{equation}
	\begin{equation}
	\Delta u(\phi)=\frac{u(\phi)-u_{min}}{u_{max}-u_{min}},
	\label{ustar}
	\end{equation}
and $u(\phi)=1/|v(\phi)|$. 
The values $\rho_{min}$, $\rho_{max}$, $u_{min}$ and $u_{max}$ are calculated over a
period of the azimuthal angle $\phi$.
When $\Delta\rho$ and $\Delta u$ vary in the same direction, $I=0$ and the
orbit is consistent. If they vary inversely, as two cosines with a phase difference 
of $\pi$, $I=\sqrt{\pi}/{2\pi}$.

\begin{figure}
\resizebox{\hsize}{!}{\includegraphics{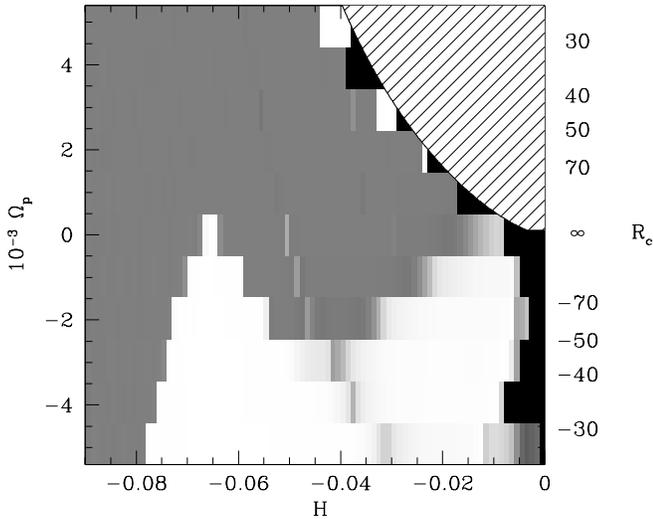}}
\caption{Consistency of orbit occupation  with the mass density as a function of the
energy and the global rotation. The dark gray corresponds to inconsistent regions while the
white are consistent regions. The black parts are missing data. 
The shading part represents the limit of the corotation. 
}
\label{zone_h}
\end{figure}

\begin{figure}
\resizebox{\hsize}{!}{\includegraphics{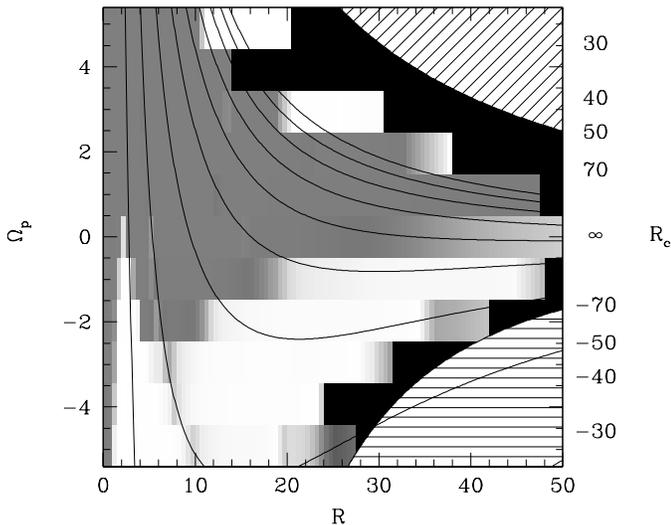}}
\caption{Same graph than in Fig.~\ref{zone_h} but as a function of radius $R$.
The upper shading part represents the limit of the corotation. While the lower 
is the region where $H>0$.
The vertical resonances ($\nu = k\, \Omega$, $k=2,\ldots 10$) are drawn in solid lines.
}
\label{zone_y2}
\end{figure}

We have tested the consistency of families \emph{pk} with respect to the energy
(Fig.~\ref{zone_h}) and radius (Fig.~\ref{zone_y2}), which is more convenient in galactic
dynamics. 
This as been calculated for different (direct and retrograde) 
pattern speeds between $-30 > R_c$ and $R_c > 30\,\textrm{kpc}$ corresponding to 
$-4.5\cdot10^{-3} < \Omega_p < 4.5\cdot10^{-3}\,\textrm{Myr}^{-1}$. The white regions 
correspond to $I=0$ (consistency) while the darker 
gray correspond to a value of $\sqrt{\pi}/{2\pi}$ (inconsistency). The shaded parts
give the limit of the computation,  
either because of the corotation or because of the positive energy regions. The black
pattern points out the missing data due to computational difficulty arising because of
the proximity of the forbidden regions. 

For a direct rotation, except in small regions near the corotation, 
the inverse local speed along an orbit varyies exactly in opposition to
the density ($I=\sqrt{\pi}/{2\pi}$). 
This causes it to depopulate the higher density regions to 
the advantage of the lower. The density distribution is also slowly modified.
However, for a slowly retrograde rotation a zone appears around $3\,\textrm{kpc}$, 
where the inverse local speed varies exactly as the density. 
This latter is also reinforced and the potential is self-consistent.  
This zone grows with increasing rotation and a second zone appears for larger radii.
For a corotation smaller than $R_c=-50\,\textrm{kpc}$ the zones of consistency cover 
the whole disk under the curve of zero energy.  

The same work applied to the subfamilies $sk\pm$ reveals that they are 
clearly inconsistent with the 
density distribution, but these families are also less relevant with the assumption 
of almost circular rotation.


\section{Conclusions}


The previous description of periodic orbits in a warped analytical potential
allows us to draw the following conclusions.

\begin{itemize}

\item[1)] The circular orbit family with an axisymmetric potential
survives the warp and becomes quasi-annular with a tilt corresponding to the
warped galactic disk. Thus, this work confirms the intuitive idea 
that the orbits follow the warp. 
Without this confirmation, the ring model of warps
would have remained baseless.
Note that the ring model has been used in the past in different contexts, but if the underlying
circular orbits do not exist, the ring model is nonphysical.

\item[2)] The existing transverse resonances in the axisymmetric case 
are also preserved in the warped case but occur at slightly lower 
energies. Subfamilies with the same period as the quasi-annular
family are generated and most of them keep their stability. In our model, the 
perturbation induced by the warp is not strong enough to generate observable 
chaotic regions.

\item[3)] Since no bifurcation with $\nu=\Omega\pm\Omega_p$ is generated by the warp, it is not possible
in our model to excite a 1:1 normal mode, which is often invoked to explain the galactic warps. 
Other modes are not favored either because the stable regions around their corresponding 
subfamilies remain very thin.

\item[4)] The study of the consistency
of the density distribution underlines the importance of the global rotation.
A direct rotation of the warp pattern
generates orbits destroying the density distribution, but on the
contrary, orbits can locally reinforce the density in the presence of
a retrograde rotation. This could help maintaining external warped
regions. 

\end{itemize}

The behavior of the discussed periodic orbits will be examined in N-body models
and the results will be the object of a future paper.


\begin{acknowledgements}
      This work as been supported by the Swiss National Science Foundation.
\end{acknowledgements}


\begin{thebibliography}{}
      
\bibitem[1997]{arnaboldi} Arnaboldi~M., Oosterloo~T., Combes~F., Freeman~K.C., Kori\-balski~B., 
      1997, Astron.~J., 113, 585A  
     
\bibitem[1990]{briggs} Briggs~F.H., 
      1990, ApJ, 352, 15     
      
\bibitem[1990]{batt} Battaner~E., Florido~E., Sanchez-Saavedra~M.L., 
      1990, A\&A, 236, 1
      
\bibitem[1998]{becq} Becquaert~J.-F., Combes~F, 
      1998, A\&A, 325, 41
      
\bibitem[1992]{binney92} Binney~J., 
      1992, ARA\&A, 30, 51      
      
\bibitem[1992]{burton} Burton~W.B., 1992, in Pfenniger~D., Bartholdi~P.,
     eds, SAAS-Fee Advanced Course 21, The Galactic Inter\-stellar Medium.
     Springer, Berlin
      
\bibitem[1991]{confwarp} Casertano~S., Sackett~P., Briggs~F.H.,
     Warped disks and inclined rings around galaxies, Cambridge University Press 1991
     
\bibitem[1980]{conto} Contopoulos~G., Papayannopoulos~T., 
     1980, A\&A, 92, 33  
     
\bibitem[1999]{debattista99} Debattista~V.P., Sellwood~J.A., 
     1999, ApJ, 513, L107  
     
\bibitem[1994]{dubi94} Dubinski~J.,
     1995, ApJ, 431, 617            
     		   
\bibitem[1995]{dubi95} Dubinski~J., Kuijken~K., 
     1995, ApJ, 442, 492
     
\bibitem[2000]{gerhard00} Gerhard~O.,
     2000, preprint, astro-ph/0010539     

\bibitem[1982]{henon} H\'enon,~M., 
     1982, Physica D5, 412
     
\bibitem[1999]{Jang99} 	Jiang~I-G., Binney~J.,
     1999, MNRAS, 303, L7     
     
\bibitem[1991]{kuijken91} Kuijken~K.,
     1991, ApJ, 376, 467  
     
\bibitem[2000]{kuijken00} Kuijken~K.,
     2000, preprint, astro-ph/0011345
                
\bibitem[1975]{nagai} Nagai~R., Miyamoto~M.,  
     1975, PASJ, 27, 533
     
\bibitem[1983]{mulder83} Mulder~W.A., 
     1983, A\&A, 121, 91

\bibitem[1984]{pfenn84} Pfenniger~D., 
     1984, A\&A, 134, 171 

\bibitem[1993]{pfenn93} Pfenniger~D., Friedli~D., 
     1993, A\&A, 270, 561 

\bibitem[1994]{pfenn94} Pfenniger~D., Combes~F., Martinet~L., 
     1994, A\&A, 285, 79
		
\bibitem[1998]{resh} Reshetnikov~V., Combes~F., 
     1998, A\&A, 337, 9
     
\bibitem[1998]{smart} Smart~R., Drimmel~R., Lattanzi~M., Binney~J.,
     1998, Nature, 392, 471 

\bibitem[1984]{spar} Sparke~L.,
     1984, MNRAS, 211, 911

\bibitem[1988]{case} Sparke~L., Casertano~S.,
     1988, MNRAS, 234, 87
     
\bibitem[1995]{weinberg95} Weinberg~M-D., 
     1995, ApJ, 455, 31
     
\bibitem[1998]{weinberg98} Weinberg~M-D., 
     1998, MNRAS, 299, 499
     
\end{thebibliography}
\end{document}